\documentclass[epsf]{hep99}
\usepackage{epsf}
\begin{document}

\title{Electroweak phase transition: recent results}

\author{F. Csikor} 

\address{ Institute for Theoretical Physics, E\"otv\"os University, H-1117 Budapest, P\'azm\'any P\'eter s\'et\'any 1A, Hungary \\[3pt]
E-mail: {\tt csikor@ludens.elte.hu}}

\abstract{
Recent results of four-dimensional (4d) lattice simulations on the finite 
temperature electroweak phase transition (EWPT) are discussed. The phase 
transition is of first order in the SU(2)-Higgs model below the end point 
Higgs mass 66.5$\pm$1.4 GeV. For larger masses a  rapid cross-over appears. 
This result completely agrees with the results of the dimensional reduction 
approach. Including the full Standard Model (SM) perturbatively the end 
point is at 72.1$\pm$1.4 GeV. Combined with recent LEP Higgs mass lower 
bounds, this excludes any EWPT in the SM. A one-loop calculation of the 
static potential makes possible a precise comparison of the lattice and 
perturbative results. Recent 4d lattice studies of the Minimal Supersymmetric 
SM (MSSM) are also mentioned.
} 

\maketitle

\section{Introduction}

The observed baryon asymmetry of the universe was eventually determined 
at the EWPT \cite{KuRS}. The understanding of this asymmetry needs a 
quantitative
description of the phase transition. Unfortunately,
the perturbative approach breaks down for the physically
allowed Higgs-boson masses (e.g. $m_H>70$ GeV) \cite{pert}.
In order to understand this nonperturbative phenomenon a
systematically controllable technique is used, namely lattice
Monte-Carlo (MC) simulations. Since merely the bosonic sector is
responsible for the bad perturbative features (due to infrared problems)
the simulations are done without the inclusion of fermions.
The first results dedicated to this questions were obtained
on 4d lattices \cite{4d}. Soon after, simulations of the
reduced model in three-dimensions were initiated, as another
approach \cite{3d}.

Recently, it became clear that for large Higgs masses the EWPT does not take place, i.e. there is an end point Higgs mass above which the first order EWPT 
goes over to a rapid crossover. Since the end point mass is smaller than the 
experimental LEP lower limit of the SM Higgs mass, baryogenesis can not be 
explained in the SM. One has to explore beyond the SM scenarios, 
the most natural choice is the MSSM.

\section{End point of the electroweak phase transition in the Standard Model}

The end point of the EWPT has been studied in the SU(2)-Higgs model in 
4d simulations \cite{label1,label2}. The effects of fermions and the U(1) 
part of the 
SM have been taken into account perturbatively.
The action in standard notation reads in case of an isotropic lattice:
\begin{eqnarray}\label{eqn1}
& S= \beta \sum
\left( 1 - {1 \over 2} {\rm Tr\,} U_{pl} \right)
-\kappa\sum
{\rm Tr\,}(\varphi^+_{x+\hat{\mu}}U_{x,\mu}\,\varphi_x)
\nonumber \\
& + \sum \left\{ {1 \over 2}{\rm Tr\,}(\varphi_x^+\varphi_x) +
\lambda \left[ {1 \over 2}{\rm Tr\,}(\varphi_x^+\varphi_x)
- 1 \right]^2 \right\} 
\end{eqnarray}

\begin{figure}
\centerline{\epsfxsize=0.90 \linewidth \epsfbox{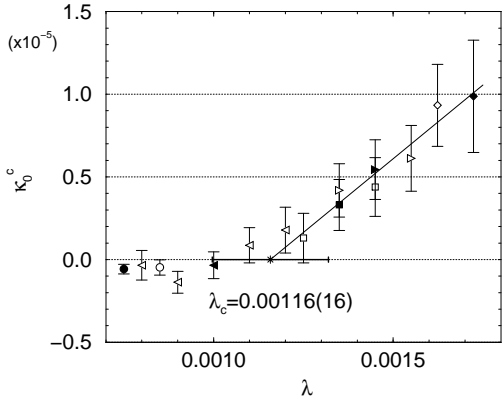}}
\caption{\label{res_lyz}
Imaginary part of first Lee-Yang zero at infinite-volume as a function 
of Higgs self coupling.}
\end{figure}

For larger time direction lattice extensions anisotropic lattices have 
been used.
The method of Lee-Yang zeros of the partition function has been applied to 
determine  the presence or absence of a first order phase transition at a 
given value of the parameters $\beta$ and $\lambda$. Fig. \ref{res_lyz} 
shows the
values of the imaginary part $\kappa_0$ of the position of the first 
Lee-Yang zeros
extrapolated to infinite (space) volume. For first order phase transitions  
the value is consistent with zero. This condition determines $\lambda_c$, which 
in turn determines (through $T$=0 simulation) the end point value 
of the Higgs mass.

\begin{figure}
\centerline{\epsfxsize=0.90 \linewidth \epsfbox{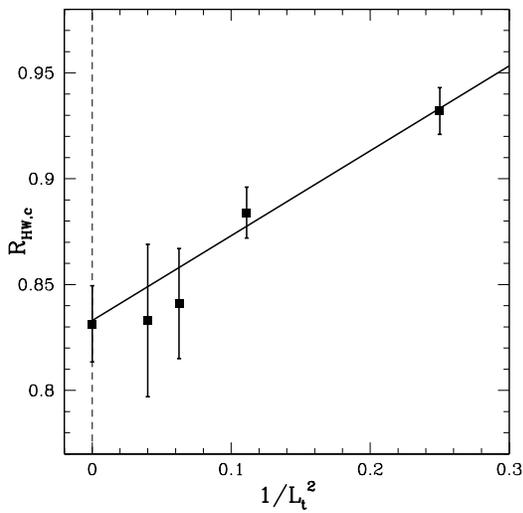}}
\caption{\label{fig2}
Dependence of $R_{HW,c}$, i.e. $R_{HW}=M_H /M_W$ corresponding to the endpoint
 of first order phase transitions on $1/L_t^2$ and extrapolation to the
  infinite volume limit.}
\end{figure}

An extrapolation to the continuum limit along the end point line of constant physics has been performed. The result is shown 
in fig. \ref{fig2}.  The critical Higgs mass is $66.5\pm 1.4$GeV in the 
SU(2)-Higgs model, in perfect agreement with the results of 3d simulations 
\cite{label5}. 

\begin{figure}
\centerline{\epsfxsize=0.90 \linewidth \epsfbox{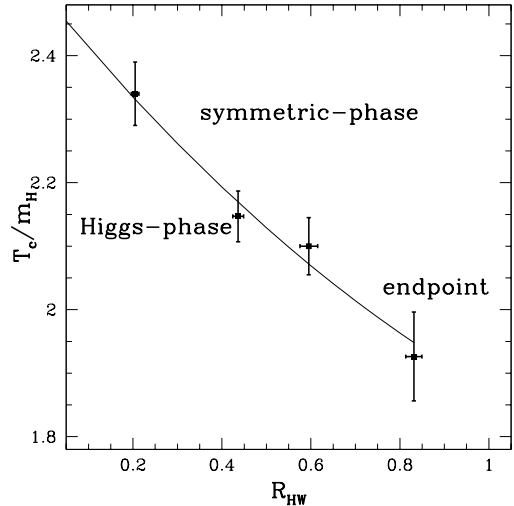}}
\caption{\label{fig3v}
{  Phase diagram of the SU(2)-Higgs model in the ($T_c /m_H - R_{HW} $)
plane. The continuous line -- representing the phase-boundary -- is a quadratic
fit
 to the data points.
   }}
   \end{figure}

Using published data of the DESY group \cite{label3}  the phase diagram can also be drawn (cf. fig. \ref{fig3v}).

\section{Renormalized gauge coupling}

 To determine the SM value of the end point Higgs mass the relation 
 of the lattice renormalized gauge coupling and the continuum 
 $\overline{\rm MS}$ coupling should be clarified. This connecion has been established in 
 \cite{label6} calculating the static potential in the continuum theory at 
 one loop level and defining the lattice analogue of the renormalized gauge coupling. The approach of the second paper in \cite{label6} has been followed  
in  correcting the end point Higgs mass of the SU(2)-Higgs model to the full to
 SM case. The result is  $M_{H_{critical}}=$72.1$\pm$1.4GeV.

\begin{figure}
\centerline{\epsfxsize=1.0  \linewidth \epsfbox{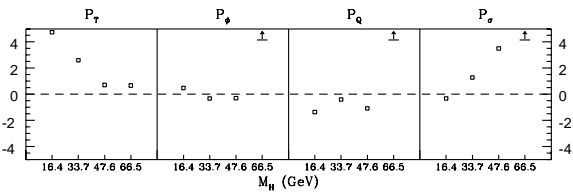}}
\caption{\label{pulls}
The pull for $T_c/m_H$, $\varphi/T_c$, $\Delta Q/T_c^4$ and
 $\sigma/T_c^3$\ as function of the Higgs mass. }
\end{figure}

 Incorporation of the precise relation between the two renormalized gauge 
 coupling definitions also allows for a better comparison of the two-loop 
 perturbative \cite{perto} and the lattice results \cite{label3}. 
The pull for the critical temperature over Higgs mass ($T _c/m_H$),
jump of the order parameter over critical temperature ($\varphi/T_c$),
normalized latent heat ($\Delta Q/T_c^4$) and normalized surface tension
($\sigma/T_c^3$) as function of the Higgs mass
 is shown in fig. \ref{pulls}. Since the perturbative approach  
does not show the end point the pulls are large for large Higgs masses.

\section{MSSM 4d simulations}

As the SM does not have a first order EWPT, the explanation of baryogenesis 
requires extended models. The MSSM has been studied perturbatively 
\cite{label7} and at two-loop order seems to yield much stronger EWPT than 
the SM. Lattice studies in a 3d reduced model also show quite a strong 
EWPT \cite {label8}.

A 4d lattice study of the bosonic part of the MSSM has been performed in 
\cite{label9}. Both Higgs doublets, the stop, sbottom and SU(2) and SU(3) 
gauge fields have been included. The simulations have been performed at 
different time extentions making possible a continuum extrapolation along a 
line of constant physics. The simulation corresponds to $\tan \beta (T=0) 
\approx 6 $ and the mass of the lighter Higgs boson is around 35 GeV.
For the physical value of $\alpha_s$ $v/T_c \approx 1.5 $, while for a 
smaller value of $\alpha_s$ a larger value was obtained.

According to the standard scenario the generated baryon asymmetry is 
proportional to \\ 
$<v^2 /T^2> \Delta \beta (T_c )$, where $v^2=v_1^2 + v_2^2 $ and 
$<v^2 >$ denotes an integral over the bubble wall and 
$\tan \beta =v_2 /v_1 $.
The $\beta $ parameter is measured in both phases and the difference turns out 
to be $\Delta \beta$=0.0045(7). This is far below the perturbative prediction 
$\Delta \beta (pert. )$=0.017.

\section{Conclusions}

The end point of hot EWPT with the technique of Lee-Yang zeros 
from simulations in 4d SU(2)-Higgs model was determined.
The phase transition is first order for Higgs masses less than
$66.5 \pm 1.4$ GeV, while for larger Higgs masses only a rapid cross-over
is expected. The phase diagram of the model was given.

It was shown non-perturbatively that for the bosonic sector of the
SM the dimensional reduction procedure works within a few
percent. This indicates that the analogous perturbative
inclusion of the fermionic sector results also in few percent error.
In the full SM we get $72.1 \pm 1.4$ GeV for the end point,
which is below the lower experimental bound. This fact is a clear
sign for physics beyond the SM.

Based on a one-loop calculation of the static potential in the
SU(2)-Higgs model a direct comparison between the perturbative
and lattice results was performed.

The MSSM is more promising for a succesfull baryogenesis. Some
4d results were shown, indicating a strong first order
phase transition.

{\bf Acknowledgments:} This work was partially supported by
Hung. Grants No.
OTKA-T22929-29803-M28413-FKFP-0128/1997.

\end{document}